\documentclass[useAMS]{mn2e}  
                                  
\usepackage{graphicx}

\newcommand{\vel}{{\rm v}}
\newcommand{\ion}[2]{{\rm #1{\small #2}}}
\newcommand{\kms}{\,{\rm km\,s}^{-1}}
\newcommand{\rev}[1]{{ #1}}
\newcommand {\mo}{\,{M}_\odot}

   \title{Stationary models for the extra-planar gas in disc galaxies}

\author[F. Marinacci et al.]{F. Marinacci\thanks{E-mail:
federico.marinacci2@unibo.it}, F. Fraternali, L. Ciotti and C. Nipoti \\\\
Dipartimento di Astronomia, Universit\`{a} di Bologna, via Ranzani 1, 40127 Bologna, Italy\\}

\begin{document}

   \date{Accepted 2009 October 2.  Received 2009 September 10; in original form 2009 June 3}

\pagerange{\pageref{firstpage}--\pageref{lastpage}} \pubyear{2009}

\maketitle

\label{firstpage}

\begin{abstract}
The kinematics of the extra-planar neutral and ionised gas in disc
galaxies shows a systematic decline of the rotational velocity with
height from the plane (vertical gradient).  This feature is not expected
for a barotropic gas, whilst it is well
reproduced by baroclinic fluid homogeneous models.  The problem with the
latter is that they require gas temperatures (above $10^5$ K) much higher
than the temperatures of the cold and warm components of the
extra-planar gas layer.  In this paper, we attempt to overcome this
problem by describing the extra-planar gas as a system of gas clouds
obeying the Jeans equations. In particular, we consider models having
the observed extra-planar gas distribution and gravitational potential
of the disc galaxy NGC\,891: for each model we construct pseudo-data
cubes and we compare them with the HI data cube of NGC\,891. In all
cases the rotational velocity gradients are in qualitative
agreement with the observations, but the synthetic and the observed data
cubes of NGC\,891 show systematic differences that cannot be accommodated
by any of the explored models.  We conclude that the extra-planar gas in
disc galaxies cannot be satisfactorily described by a stationary
Jeans-like system of gas clouds.
\end{abstract}


%

\section{Introduction}

Observations of several spiral galaxies at various wavelengths have
revealed massive gaseous haloes surrounding the galactic discs.  This
extra-planar gas is multiphase: it is detected in \ion{H}{I} (e.g. Swaters, Sancisi
\& van der Hulst 1997), in H$\alpha$ (Rand 2000; Rossa et al. 2004) and in X-ray
observations (Wang et al. 2001; Strickland et al. 2004). Thanks to
high-sensitivity \ion{H}{I} observations of some edge-on galaxies, like NGC
891 (Oosterloo, Fraternali \& Sancisi 2007), we can trace these haloes up to $10$ -
$20$ kpc from the plane and we can study their kinematics in
detail.
A remarkable feature of the extra-planar gas is its
regularly decreasing rotational velocity $u_{\varphi}$ at increasing
distances from the galactic plane (vertical gradient). 
Fraternali et al.\ (2005) found that
in NGC 891 the vertical gradient is $\sim\, -15 {\rm km \,s} ^{-1}
{\rm kpc}^{-1}$ (see also Heald, Rand \& Benjamin et al. 2007). For other galaxies (e.g. NGC 2403) gradients of the
same order have been estimated (see Fraternali 2008).

The dynamical state of the extra-planar gas can provide useful
insights to understand its origin.  Two main frameworks have been
proposed: the galactic fountain (Shapiro \& Field 1976; Bregman 1980)
and the cosmological accretion (Oort 1970; Binney 2005).  In the
galactic fountain model, partially ionized gas is ejected from the
disc by supernova explosions and stellar winds, it travels through the
halo and eventually falls back to the disc. Due to difficulties
in running high-resolution hydrodynamical simulations of the whole
galactic disc (see e.g.\ Melioli et al.\ 2008), the gas in the
galactic fountain model is often treated ballistically, i.e. as a
collection of non-interacting clouds subject to the gravitational
field of the galaxy (Collins, Benjamin \& Rand 2002). The ballistic galactic fountain is able to
reproduce the gas distribution observed in spiral galaxies, but in
general the predicted vertical gradient of $u_{\varphi}$ is
significantly lower than observed (Fraternali \& Binney 2006).  In the
accretion models, the extra-planar gas is the result \rev{of the infall of cool intergalactic gas}.  
The two scenarios are not mutually exclusive: in fact it has been argued that the
contribution of accretion is limited to only $10-20\%$ of the total
mass of extra-planar gas (Fraternali \& Binney 2008).

In a complementary approach, some authors (Benjamin 2002; Barnab\`{e}
et al. 2006), have focused on the construction of models for the
extra-planar gas under the assumption of stationarity. In this approach 
the problem of the origin of the extra-planar gas is not addressed, and the 
main effort is to clarify the global dynamical state of the gas. Usually, in the stationary
approach, the
extra-planar gas is described as an homogeneous fluid in permanent
rotation. In the simplest of fluid homogeneous models,
vertical and radial motions of the gas are neglected and
therefore the vertical gravitational field of the galaxy is balanced
only by the pressure gradient of the gas. In other studies, turbulence has also been added as an extra-pressure term
(e.g., Koyama \& Ostriker 2008). According to the
Poincar\'{e}-Wavre theorem (Lebovitz 1967; Tassoul 1978), a vertical
gradient in the rotational velocity can be present only if the
pressure of the medium is not stratified on the density, therefore in fluid homogeneous models the
gas distribution is necessarily \textit{baroclinic} (Barnab\`{e} et al. 2006; see
also Waxman 1978). Remarkably, in their study of NGC 891, Barnab\`{e} et al. (2006) found that if
the baroclinic configuration is built from general physical arguments,
the model rotational velocity well reproduces the observed rotation curves
at different heights over the galactic disc and for a large radial range.
However, the temperature predicted for the system is $> 10^{5}$ K, well above
that of the neutral gas.

Here we investigate if and how this ``temperature problem'' can be solved.
We make use of the fact that stationary fluid equations for a gaseous system in
permanent rotation are, from a formal point of view, identical to the
stationary Jeans equations for an axisymmetric system with isotropic
velocity dispersion. Thus, also in this case, a decrease of
rotational velocities with increasing distances from the mid-plane is
obtained when the density field is built using the approach described
by Barnab\`{e} et al. (2006). In this paper we analyse this
alternative interpretation of the baroclinic fluid homogeneous
models. 
In practice we consider a ``gas'' of cold \ion{H}{I} clouds described by the
stationary Jeans equations, where the thermal pressure, needed for the
vertical balance of the galaxy gravitational field, is replaced by the
velocity dispersion of the clouds.

\rev{
In the present approach, some preliminary consideration on the physical state
of the clouds is important.
In fact, \ion{H}{I} halo clouds are often assumed to be
embedded in, and in pressure equilibrium with, a hot medium (corona) 
that is about a thousand times less dense and provides the pressure required to 
confine them (Spitzer 1956; Wolfire et al. 1995).
In the Milky Way, there is plenty of evidence for the existence of this 
medium, such as emission from highly ionised metals (e.g. Sembach et al.\ 2003) 
and the head-tail morphology of individual High Velocity Clouds (HVCs,
e.g. Br{\"u}ns et al. 2000).
Due to the high density contrast, the pressure-confined clouds 
cannot be supported by the pressure of the external medium and must be moving 
more or less ballistically in a way akin to water drops in the air
(Bregman 1980).
This assumption is commonly employed in galactic fountain models 
(e.g. Collins et al. 2002). 
On the time-scale taken for a cloud to move about 1000 times 
its length, the trajectory may deviate significantly from the ballistic one due
to the interactions with the external medium.
This time-scale depends on the cloud mass but it is
longer than a dynamical time for the larger clouds (Fraternali \& Binney 2008).
The typical mass of an halo cloud is difficult to estimate. 
In the Milky Way the new determinations of distances for the HVCs give 
masses from a few $\times 10^4 \mo$ to $10^6 \mo$ (e.g. Wakker et al. 2008), 
for external galaxies the 
mass resolution is often above $10^6 \mo$ but in M31 several clouds with
masses down to $\sim 1 \times 10^5 \mo$ have been observed
(Thilker et al. 2004).
We estimate that for a cloud of $10^5 \mo$, the drag time scale is more than
10 times the dynamical time and it is therefore fair to 
treat the system as purely dynamical.
Moreover, if the \ion{H}{I} halo of a galaxy like NGC\,891 is made up of clouds
with this mass and radii of $100$ pc, the collision time between clouds
turns out to be about 5 times the dynamical time and the system can be
considered collision-less.
}

The substitution of a fluid system with
a gas of clouds is also not trivial from the point of view of the comparison with
the observations, as several delicate issues arise when considering \ion{H}{I}
observations (see Sect. \ref{constr}).
To overcome these problems we follow Fraternali \&
Binney (2006, 2008) and construct, as an output for each of the models
investigated, a \textit{pseudo-data cube} with the same resolution and
total flux as the \textit{data cube} of the \ion{H}{I} observations.
This procedure assures a full control of projection and resolution
effects. Moreover, the comparison with the raw data removes all the
intermediate stages of data analysis and the associated uncertainties.
For completeness, we also investigate some phenomenological anisotropic models.

The paper is organized as follows: in Section 2 we illustrate the
Jeans-based interpretation of the baroclinic solutions,
introducing 
the anisotropic case which has no analogous fluid counterpart, and we
discuss the conditions required for these solutions to have a negative
vertical velocity gradient. In Section~3 we present the method adopted
to construct isotropic and anisotropic models for the edge-on galaxy
NGC 891, whilst in Section~4 we compare their predictions with the
\ion{H}{I} observations. Section~5 is devoted to the discussion of the
results, and Section~6 concludes.

\section{Jeans-based description of fluid baroclinic solutions}

\subsection{Isotropic Jeans solutions}
\label{sect:iso}

Consider an axisymmetric density distribution of gas clouds moving
under the influence of an axisymmetric galactic gravitational
potential $\Phi (R,z)$, so that all the physical quantities are
independent of the azimuthal angle $\varphi$ of cylindrical
coordinates. The galactic gravitational potential $\Phi$ is the sum of
the dark-matter halo, the bulge and the stellar and gaseous disc
potentials. We neglect the contribution to the gravitational field
of the extra-planar gas, which represents a very small fraction of
the total mass (for instance, less than $1\%$ in NGC 891). If the
velocity dispersion tensor of the cloud distribution is isotropic the
associated stationary Jeans equations are
    \begin{equation}
	\label{isojeans}
     \left\{
     \begin{array}{l}
         \displaystyle \frac{1}{\rho}\frac{\partial\rho\sigma_{z}^{2}}{\partial z} = -\displaystyle \frac{\partial \Phi}{\partial z} \,, \\
         \\
         \displaystyle\frac{1}{\rho}\frac{\partial \rho\sigma_{R}^{2}}{\partial R} = -\displaystyle \frac{\partial\Phi}{\partial R} + \frac{\overline{\vel_{\varphi}^{2}}-\sigma_{R}^{2}}{R} \,,
      \end{array}
      \right.
    \end{equation}
where $\rho$ is the cloud density distribution, $\sigma_{R} =
\sigma_{z} = \sigma$ is the one-dimensional velocity dispersion, and
$\vel_{\varphi}$ is the velocity of each cloud around the $z-$axis
(e.g., Binney \& Tremaine 2008). The phase-space average is indicated
by a bar over the quantity of interest so $u_{\varphi} =
\overline{\vel}_{\varphi}$, and we are assuming that $u_{R} = u_{z} =
0$, i.e. no net motion in the radial and the vertical directions is
considered. Under these assumptions the streaming velocity of the
clouds is given by

    \begin{equation}
	\label{rotvel}
         \displaystyle u_{\varphi}^{2} = \overline{\vel_{\varphi}^{2}} - \sigma^{2} \,,
     \end{equation}
and eqs. (\ref{isojeans}) are formally identical to the equations
describing a stationary fluid in permanent rotation, where the thermodynamic
pressure of the fluid is replaced by the quantity $\rho\sigma^{2}$.

To solve eqs. (\ref{isojeans}) we fix the galactic
gravitational potential $\Phi(R,z)$ and we assume a cloud density
distribution $\rho(R,z)$ consistent with the observations and vanishing at infinity (see Section \ref{sect:application}). With this procedure, the obtained configuration is in general baroclinic. Therefore, from the
first of eqs. (\ref{isojeans}) with  boundary condition
$\rho\sigma_{z}^{2} \to 0$ for $z \to \infty$, one obtains

    \begin{equation}
	\label{pressure}
         \displaystyle \rho\sigma_{z}^{2} =  \displaystyle \int_{z}^{\infty}\rho\frac{\partial \Phi}{\partial z} \, {\rm d}z \,,
    \end{equation}
    and $u_{\varphi}$ is calculated from the second of the
    eqs. (\ref{isojeans}) and eq. (\ref{rotvel}). Note that the rotational velocity field
    $u_{\varphi}$ can be obtained in general by the  ``commutator-like''
    relation
     \begin{equation}
	\label{isovelocity}
         \begin{array}{ll}
         \displaystyle \frac{\rho u_{\varphi}^{2}}{R} =  \displaystyle \int_{z}^{\infty}\left(\frac{\partial \rho}{\partial R}\frac{\partial \Phi}{\partial z} - \frac{\partial \rho}{\partial z}\frac{\partial\Phi}{\partial R} \right) \, {\rm d}z \,,
      \end{array}
    \end{equation}
(e.g., see Barnab\`{e} et al. 2006; and references therein).

The expression above reveals that the existence of physically
acceptable solutions with $u_{\varphi}^{2} \geq 0$ everywhere is not
guaranteed for independent choices of $\rho$ and $\Phi$. In
Barnab\`{e} et al. (2006) general rules for
constructing physically acceptable baroclinic solutions are
illustrated. For example, in the usual case in which $\partial \Phi /
\partial R \geq 0$ and $\partial \Phi / \partial z \geq 0$, a
sufficient condition to ensure the positivity of $u_{\varphi}^{2}$ is
that $\partial \rho / \partial z \leq 0$ and $\partial \rho / \partial
R \geq 0$.  The same rules apply also here suggesting the use of
centrally depressed density distributions. This implies that in the
central regions the density is radially increasing moving outwards,
which is consistent with the radial \ion{H}{I} distribution observed
in spiral galaxies such as NGC 891 (Oosterloo et al. 2007).

\subsection{Anisotropic Jeans solutions}
\label{sect:aniso}

In Section \ref{sect:application} we will show that Jeans isotropic
models, \rev{the natural kinetic counterpart of homogeneous fluid models},
fail to reproduce some features of the extra-planar gas in NGC
891. For this reason we also explore a phenomenological model
including velocity dispersion anisotropy.  Given the
gravitational potential $\Phi$ and the cloud density distribution
$\rho$, the solution of the stationary axisymmetric Jeans equations
with an anisotropic velocity dispersion tensor is unique if one
assigns the orientation and the shape of the intersection of the
velocity dispersion ellipsoid everywhere in the meridional plane. For
our problem, this is done by maintaining $u_{R} = u_{z} = 0$ and
assuming a suitable parametrization for the anisotropy, which
specifies the ratio $\sigma_{R}/\sigma_{z}$.  The simplest choice is
that of a constant anisotropy which, following Cappellari (2008), we
express as $\sigma_{R}^{2} = b\sigma_{z}^{2}$ with $b \geq 0$ \footnote{Note
that this anisotropy cannot be realized (except for $b = 1$) under the
common assumption of a two-integral distribution function depending on
energy and angular momentum along the $z-$axis, as in this case
$\sigma_{R} = \sigma_{z}$.}.  Thus, eqs. (\ref{isojeans}) become

    \begin{equation}
	\label{anisojeans}
     \left\{
     \begin{array}{ll}
         \displaystyle \frac{1}{\rho}\frac{\partial\rho\sigma_{z}^{2}}{\partial z} & = -\displaystyle \frac{\partial \Phi}{\partial z} \,,\\
         \\
         \displaystyle\frac{b}{\rho}\frac{\partial \rho\sigma_{z}^{2}}{\partial R} & = -\displaystyle \frac{\partial\Phi}{\partial R} + \frac{\overline{\vel_{\varphi}^{2}}-b\sigma_{z}^{2}}{R} \,.
      \end{array}
      \right.
    \end{equation}

Thus $\sigma_{z}^{2}$ is given by eq.~(\ref{pressure}) as in the
  isotropic case, whilst it can be shown that
     \begin{equation}
	\label{anisovelocity}
         \begin{array}{lll}
         \displaystyle \frac{\rho(\overline{\vel_{\varphi}^{2}} - b\sigma_{z}^{2})}{R} = \\ \\  \qquad\qquad  \displaystyle b\int_{z}^{\infty}\left(\frac{\partial \rho}{\partial R}\frac{\partial \Phi}{\partial z}-\frac{\partial \rho}{\partial z}\frac{\partial\Phi}{\partial R} \right) \, {\rm d}z \, + (1 - b)\, \rho\displaystyle\frac{\partial \Phi}{\partial R} \,. 
      \end{array}
    \end{equation}
Note that equation above can be rewritten in a more compact form as
     \begin{equation}
	\label{anisovelocity2}
         \displaystyle \overline{\vel_{\varphi}^{2}} - b\sigma_{z}^{2} =   \displaystyle b u_{\rm iso}^{2} + (1 - b)\, u_{\rm c}^{2}  \,, 
    \end{equation}
where $u_{\rm iso}$ is the rotational velocity field of the
corresponding isotropic model (defined as the model with the same
potential and the same density distribution, but with $k = b = 1$)
and
    \begin{equation}
	\label{vcirc}
         \displaystyle u_{\rm c}^{2} \equiv R\, \frac{\partial \Phi}{\partial R} 
    \end{equation}
is the circular velocity at position $(R, z)$.
The streaming velocity field is obtained by splitting $\overline{\vel_{\varphi}^{2}}$ into the streaming motion $u_{\varphi}^{2}$ and the azimuthal velocity dispersion $\sigma_{\varphi}^{2}$. Following the widely used Satoh (1980) $k$-decomposition we have

    \begin{equation}
     \label{satohdecomp}
         \displaystyle u_{\varphi}^{2} = k^{2}(\overline{\vel_{\varphi}^{2}} - b\sigma_{z}^{2}) \,,\\
    \end{equation}
so that
    \begin{equation}
     \label{satohdecomp2}
         \displaystyle\sigma_{\varphi}^{2} = b\sigma_{z}^{2} + (1 - k^{2})\,(\overline{\vel_{\varphi}^{2}} - b\sigma_{z}^{2}) \,,
     \end{equation}
with $0 \leq k \leq 1$; in general $k$ can be function of the
position, i.e. $k = k(R, z)$ (e.g., see Ciotti \& Pellegrini
1996). For $k = 0$ no ordered motions are present, whilst
$\sigma_{\varphi}^{2} = \sigma_{R}^{2}$ in the case $k = 1$; the
isotropic case is recovered for $k = b = 1$.  From
eq. (\ref{satohdecomp}) it follows that the Satoh decomposition
procedure can be applied only when $\overline{\vel_{\varphi}^{2}} -
b\sigma_{z}^{2} \geq 0$, so $k$ and $b$ are not fully
independent. Note that if the corresponding isotropic model leads to
a physically acceptable solution, then a sufficient condition to
ensure the positivity of $u^{2}_{\varphi}$ is $b \leq 1$: if
$u^{2}_{\rm iso} \geq 0$, the l.h.s. of eq.~(\ref{anisovelocity2}) can
be negative only if $b > 1$, so the use of the Satoh decomposition is
consistent with the assumption $\sigma_{R} \leq \sigma_{z}$.  This
requirement is in agreement with the observations that show large
vertical motions of the order of $50-100 \kms$ and velocity dispersion
along $R$ and $\varphi$ of the extra-planar gas of approximately $20
\,{\rm km \, s^{-1}}$ (Boomsma et al. 2008; Oosterloo et
al. 2007). Summarizing, this simple parametrization of the velocity
dispersion anisotropy is consistent with physically acceptable
solutions provided that: (i) $b \leq 1$ and (ii) the corresponding
isotropic model has $u_{\varphi}^{2} \geq 0$ everywhere.

\subsection{Conditions to have negative vertical gradient}

An important property of the extra-planar gas is the decrease of
its rotational velocity with the distance from the galactic plane.  To
study the kinematics of the extra-planar gas in the models it is
helpful to have general expressions (such as eq. [2] of Waxman 1978)
relating the vertical gradient of the rotational velocity to other
physical quantities in stationary configurations.

For example, in the isotropic case we differentiate
eq. (\ref{isovelocity}) with respect to $z$ and obtain
\begin{equation}
	\label{verticalgradient}
         \displaystyle \displaystyle\frac{\partial u_{\rm iso}^{2}}{\partial z} =  \displaystyle \frac{\partial \ln\rho}{\partial z}\displaystyle\left(u_{\rm c}^{2} - u_{\rm iso}^{2} \right) - \frac{\partial \ln\rho}{\partial \ln R}\frac{\partial \Phi}{\partial z} \,.
\end{equation}
Thus, the necessary and sufficient condition to have a decrease in
the rotational velocity in the isotropic model is the negativity of
the right hand side of eq. (\ref{verticalgradient}). In particular,
consider a typical potential such that $\partial \Phi / \partial z
\geq 0$ and a centrally depressed density distribution vanishing
for $R\to\infty$ with $\partial \rho / \partial z \leq 0$.  For each
value of $z$ there is a radius $R_{\rm max}$ such that $\rho$
increases with $R$ for $R<R_{\rm max}$ and then decreases with $R$ for
$R \geq R_{\rm max}$.  From eq. (\ref{verticalgradient}) it follows
that for $R \geq R_{\rm max}$ a necessary condition for a solution to
exhibit a negative vertical gradient in the rotational velocity is
$u_{\rm iso}^{2} \leq u_{\rm c}^{2}$.

For systems with anisotropic velocity dispersion tensor of the class here considered, there is not a
simple expression for the vertical gradient of the rotational
velocity, unless we assume that $k$ is independent of $z$.
In this case, a useful expression for the vertical gradient of the
rotational velocity can be obtained by substituting
eq. (\ref{satohdecomp}) into eq. (\ref{anisovelocity2}) and by
differentiating the result with respect to $z$. This leads to

    \begin{equation}
	\label{anisovertgrad}
         \displaystyle \frac{\partial \,u_{\varphi}^{2}}{\partial z} = k^{2}\left[\displaystyle b\frac{\partial\,u_{\rm iso}^{2}}{\partial z} + (1 - b)\frac{\partial \,u_{\rm c}^{2}}{\partial z}\right] \,,
    \end{equation}
    where $u_{\rm iso}$ is the rotational velocity of the
    corresponding isotropic model. It follows that in the usual cases,
      in which $\partial u_{\rm c}^{2}/\partial z \leq 0$ and $b < 1$ (see Section \ref{sect:aniso}), a negative vertical velocity gradient
    in the corresponding isotropic model is a sufficient
    condition to have $\partial u_{\varphi} /\partial z \leq 0$ also in the anisotropic models of the family here considered.

\section{Application to NGC 891}
\label{sect:application}

The general rules of Section 2 are now applied to the specific case of
the extra-planar gas of NGC 891, to explore whether the proposed
models reproduce the observed decrease of the rotational velocity for
increasing $z$.  We note that the resulting models can directly
describe the extra-planar \ion{H}{I} gas seen in the observations
because the vertical ``pressure'' support, needed against the galaxy
gravitational field, is now provided by the velocity dispersion
of the clouds, rather than by the thermodynamic temperature as in the
homogeneous baroclinic models of Barnab\`{e} et al. (2006). The
present investigation differs from that of Barnab\`{e} et al. (2006)
also because we adopt an observationally justified density
distribution for the extra-planar gas and we perform the comparison of
the models with the observational data by constructing a pseudo-data
cube for each model with the same resolution and total flux as the
observations.

\begin{figure}
\hspace{-0.5cm}
\includegraphics[width=10cm,height=7cm,viewport=19 90 450 330]{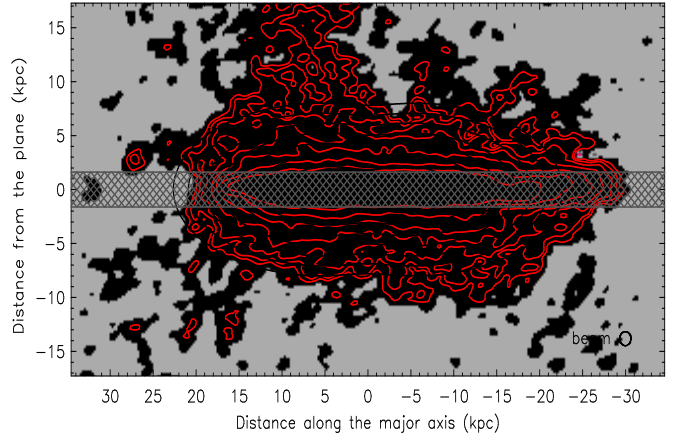}
\caption{Total \ion{H}{I} emission map of NGC 891 as obtained by the
  observations (red contours and colour scale) and by the cloud
  density distribution of Section \ref{cloud distr} (black
  contours). The map has a spatial resolution of 25\arcsec ($\sim1.2$
  kpc) with contour levels, starting from the outermost, of 0.15,
  0.3, 0.6, 1.2, 2.4, 4.8, 9.6 $\times 10^{20} \, \rm{cm}^{-2}$. The
  shaded area delimits the emission expected from the disc.}
\label{fig:map}
\end{figure}

\subsection{The galaxy model}

For the determination of the gravitational potential of NGC 891 we
consider the maximum-light mass model of Fraternali \& Binney
(2006). This mass model consists of four distinct components: two
exponential discs (stellar and gaseous), a bulge and a dark-matter
  halo. The volume density of the two disc components is given by
	\begin{equation}
         \displaystyle \rho(R, z) =  \displaystyle \rho_{0}{\rm e}^{-R/R_{\rm d}} \,{\rm sech}^{2}\left(\frac{z}{h_{\rm d}}\right) \,,
	\end{equation}
where $\rho_{0}$ is the central density, and the dimensional
constant $R_{d}$ and $h_{d}$ are, respectively, the scale-length and
the scale-height of the disc. The bulge and the dark halo components
are modelled using the double power-law spheroidal density profile
(Dehnen \& Binney 1998)
	\begin{equation}
         \displaystyle \rho(m) =  \displaystyle \frac{\rho_{0}}{(m/a)^{\alpha}\,(1 + m/a)^{\beta - \alpha}}\,,
	\end{equation}
where
        \begin{equation}
         \displaystyle m^{2} =  \displaystyle R^{2} + \frac{z^{2}}{1 - e^{2}} \,,
	\end{equation}
$e$ is the eccentricity of the spheroid, $\alpha$ and $\beta$ are the
inner and outer slopes, and $a$ is the scale-length. The adopted
values of the parameters to reproduce the rotation curve in the plane of NGC 891
are reported in Table 1 of Fraternali \& Binney (2006), where the
derivation of this potential and other mass models compatible with this 
rotation curve are also discussed.

\subsection{The cloud distribution}
\label{cloud distr}

A proper Jeans-based analysis of baroclinic fluid solutions requires
the choice of a cloud density distribution similar to that observed. For 
the cloud density distribution we adopt the function
\begin{equation}
  \label{densitydistrib}
  \displaystyle \rho(R, z) =  \displaystyle \rho_{0}\left(1 + \frac{R}{R_{{\rm g}}}\right)^{\gamma} {\rm e}^{-R/R_{{\rm g}}} \,{\rm sech}^{2}\left(\frac{z}{h_{\rm g}}\right)\,,
\end{equation}
which is similar to that adopted by Oosterloo et al. (2007). In the
equation above $\rho_{0}$ is the central density, $\gamma \geq 0$,
$R_{{\rm g}}$ is the scale-length and the scale-height $h_{g}$ depends
upon $R$ as
	\begin{equation}
	\label{gasheight}
         \displaystyle   h_{\rm g}(R) = h_{0} + \displaystyle \left(\frac{R}{h_{R}}\right)^{\delta},\;\, \delta \geq 0.
	\end{equation}
The parameters of the distribution, excluded
the central density $\rho_{0}$, are chosen is such a way that the
function reproduces the \ion{H}{I} total map shown in
Fig. \ref{fig:map} and their values are: $R_{{\rm g}} = 1.61$ kpc,
$\gamma = 4.87$, $h_{0} = 2.3$ kpc, $\delta = 2.25$ and $h_{R} =
18.03$ kpc. The central density $\rho_{0}=6.63\times 10^{-4} {\rm
cm}^{-3}$ is such that the total mass of the cloud system is that
derived by the total \ion{H}{I} emission; the corresponding halo mass
(i.e.\ the mass computed for $|z| \geq 1$ kpc) of $1.13 \times
10^{9}{\rm M}\odot$ is consistent with the value given in Oosterloo et
al.\ (2007).

We now verify that the density distribution guarantees the
existence of physically acceptable solutions (i.e. $u_{\varphi}^{2}
\geq 0$ everywhere). It can be easily shown that $\partial \rho /
\partial z \leq 0$ for the density distribution
(\ref{densitydistrib}), whilst the radial density profile presents,
for each $z$, a maximum at some value $R_{\rm max}(z)$: the positivity
of $u_{\varphi}^{2}$ is then guaranteed in the region $R \leq R_{\rm
max}$, where $\partial \rho / \partial R > 0$, whilst it has to be
investigated in the external region, where $\partial \rho
/ \partial R < 0$ (see Section \ref{sect:iso}), by numerical inspection. We found that, with
the adopted parameters, the density distribution
(\ref{densitydistrib}) produces physically acceptable solutions for
all the models investigated. For what concerns the vertical velocity
gradient, the considerations made in Section 2.3 are valid: in particular, we
expect that the rotational velocities of the models are lower than the
circular velocity (\ref{vcirc}) for $R \geq R_{\rm max}$, if the
vertical velocity gradient is negative.

\begin{figure*}
\hspace{-0.7cm}
\centering
\includegraphics[width = 17 cm,viewport=38 50 431 302]{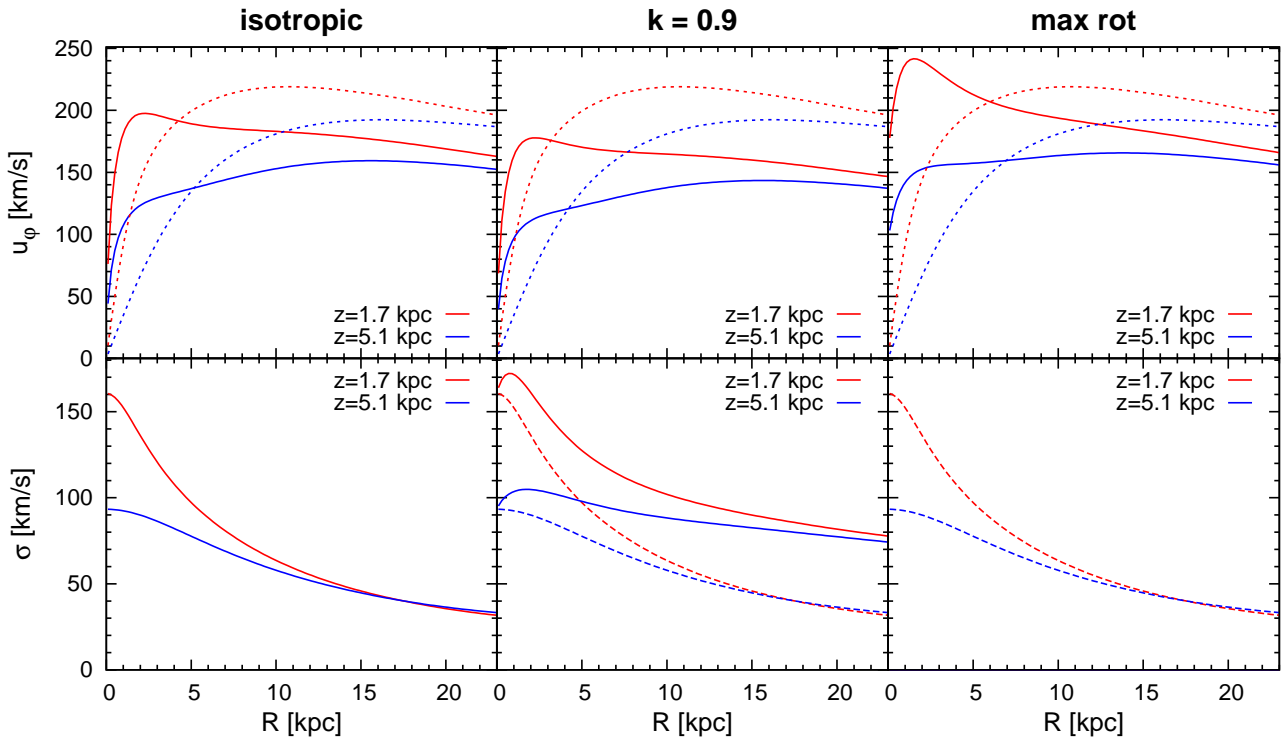}
\caption{Intrinsic rotational
  velocities and velocity dispersion for system of clouds that can be described by a classical
  two-integral phase-space distribution function, at $z = 1.7$ kpc
  and $z = 5.1$ kpc. From left to right:
  isotropic model, $k = 0.9$ model and maximally rotating model
  ($\sigma_{\varphi} = 0$). Velocity panels: the dotted lines
  represent the circular velocity $u_{\rm c}$ whilst the solid lines
  are the model rotational velocities $u_{\varphi}$. Dispersion
  panels: the dashed lines indicate $\sigma_{z}$ and the solid lines
   $\sigma_{\varphi}$. In the isotropic case all the components of the
  dispersion are the same, whilst in the other two models $\sigma_{z}
  = \sigma_{R}$.}
\label{fig:iso}
\end{figure*}

In our models, we are only aiming to reproduce the extra-planar gas
and we therefore neglect the region where emission from the disc is
expected.  Such a region is covered by shaded areas in all the figures
where the comparison with the data is shown (Figs.\ \ref{fig:map},
\ref{fig:isocube}, \ref{fig:anisocube}, and \ref{fig:v-z_plot}).  The
size of this region ($|z| < 1.64$ kpc), is calculated as $\pm
3\sigma\, \times$ (spatial resolution of the data) assuming that the
galactic disc \c{is perfectly co-planar.}  In fact the resolution of
the observations (HPBW$ = 0.64$ kpc) is much larger than the expected
intrinsic scale height of the \ion{H}{I} disc ($100-200$ pc), i.e.\
the disc is not resolved.  The exclusion of the disc is necessary to
avoid the complication of a region where the gas is almost certainly
not described by the considered Jeans equations. The implicit
assumption is that, in a real galaxy, at the lower halo boundaries the
velocity dispersion, induced by some (unspecified) physical
process in the disc, is the same as the one provided by our density
distribution (see Figs.\ \ref{fig:iso} and \ref{fig:aniso}).  This
requirement is necessary for the system to be stationary.

\subsection{Construction of the model}
\label{constr}

Having fixed the galaxy gravitational field and the cloud density
distribution, for each model we first compute $\sigma_{z}$ by
integrating numerically eq.~(\ref{pressure}) with a
finite-difference scheme.  In the isotropic case we then compute the
rotational field from eq. (\ref{isovelocity}) using the same numerical
scheme. In the anisotropic cases we can integrate
eq. (\ref{anisovelocity}) or just use (\ref{anisovelocity2}). The
rotation velocity and the azimuthal component of the velocity
dispersion are then evaluated on the numerical grid by means of
eqs. (\ref{satohdecomp}) and (\ref{satohdecomp2}). We verified that
eqs. (\ref{anisovelocity}) and (\ref{anisovelocity2}) give the same
results in all explored anisotropic models.

As an output for the models a pseudo-data cube, with the same spectral
and spatial resolutions and total flux as the data cube of \ion{H}{I}
observations, is generated as follows. In each grid point the cloud
density, the rotational velocity and the velocity dispersion are
projected numerically along the line-of-sight for the inclination of
the galaxy, which is assumed to be seen exactly edge-on. From the
projected velocity we construct a first data cube and, to account for
the effect of the velocity dispersion, we convolve the projected
density in each cell of this cube with a normalised Gaussian profile
with a dispersion equal to the projected velocity dispersion. Finally,
for a more realistic comparison, the cube is smoothed to the same
spatial resolution as the data ($FPBW = 28^{''} \approx 1.3$ kpc).  We
remind that the construction of a pseudo-data cube is crucial because
it bypasses all the intermediate stages of data analysis, and the
assumptions made in these stages.  A critical case is the comparison
of the data rotation curves with the rotational velocities predicted
by the models.  The rotation curves of the extra-planar gas (and of
edge-on galaxies in general), are usually derived by fixing a certain
amount of line-of-sight velocity dispersion and using the so-called
``envelope tracing'' method (see Sancisi \& Allen 1979;
Fraternali et al. 2005).  Clearly, if a model has velocity dispersions
inconsistent with that adopted to derive the rotation curves from the
data (as it happens here, see Section \ref{sect:results}), a
direct comparison between the two cannot be done (see Section
\ref{sect:discussion}).

\section{Results}
\label{sect:results}

We now present the most important features of some of the
considered stationary models for the extra-planar gas of NGC 891, and
we compare them with the observations.

\subsection{Isotropic model}

We start by describing the main intrinsic features of the isotropic
model, i.e. the analog of a stationary baroclinic fluid model where
the thermal pressure of the gas is replaced by a globally isotropic
velocity dispersion tensor. It is then expected that the behaviour of
the isotropic model is similar to the baroclinic solutions of
Barnab\`{e} et al. (2006). However, there is an important difference
between these two models in the choice of the density distribution.
In our model the density is chosen to match the observed \ion{H}{I}
emission (see Fig.\ \ref{fig:map}), whilst the density distribution of
Barnab\`{e} et al. (2006) was just chosen to satisfy the general rules
that guarantee a physically acceptable rotational velocity field. In
other words, our density distribution is fixed by the \ion{H}{I}
observations, and not by the requirement $u_{\varphi}^{2} \geq 0$.
Using this cloud distribution the resulting
$\sigma_{z}=\sigma_{R}=\sigma_{\varphi}$ has values ranging from 160
(in the centre) to $35\, {\rm km \, s^{-1}}$ (at $R\sim 20$ kpc) with
mean values between 50 and $100\, {\rm km \, s^{-1}}$ (Fig.\
\ref{fig:iso}, bottom left panel).

\begin{figure*}
\centering
\includegraphics[width = 17 cm,viewport=19 121 580 763]{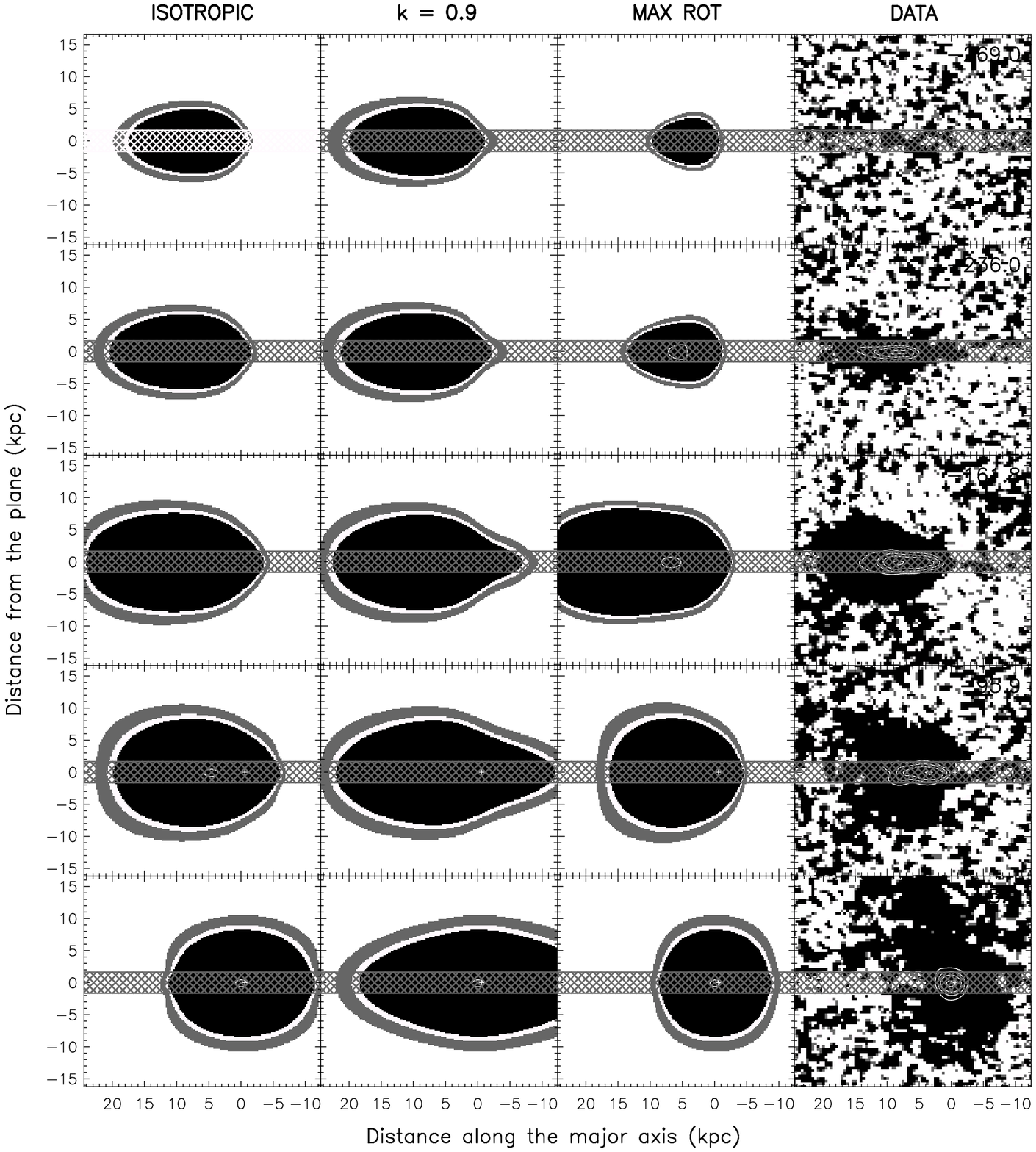}
\caption{Comparison of representative velocity channels of the
  \ion{H}{I} data cube of NGC 891 and the predictions of 3 stationary
  models: the isotropic (first column), the $k = 0.9$ and $b = 1$
  (second column), and the $b = 1$ maximally rotating (third
  column) models.  The shaded area in each channel delimits the
  emission expected from the disc.  The contour levels, starting from
  the outer, are 0.25, 0.5, 1, 2, 5, 10, 20, 50 mJy/beam; the beam
  size is 28\arcsec = 1.3 kpc.}
\label{fig:isocube}
\end{figure*}

In the left panels of Fig. \ref{fig:iso} we present the rotation curves (top)
and the radial velocity dispersion profiles (bottom) at
distances from the plane $z = 1.7$ kpc (lower halo boundary) and $z =
5.1$ kpc; the dashed lines show the circular velocities $u_{\rm
c}$. As expected from the considerations of Section 2.3, in the
external regions of the galaxy the rotational velocity is below the
circular velocity because of the negative vertical gradient in
$u_{\varphi}$. The radial profile of the velocity dispersion shows a
monotonic decrease with $R$ but the values of the dispersion are
everywhere greater than the mean observed value of $\sim 20\, \rm{km\,
s^{-1}}$.  This latter is also the value adopted to derive the
rotation curves above the plane from the \ion{H}{I} observations of
NGC\,891 (Fraternali et al. 2005, here shown in
Fig. \ref{fig:rotcurve}). As a consequence, a direct comparison
between these curves and the circular velocities of our models is not
possible.  We stress again that these problems are overcome by
comparing directly the models with the raw data cubes, where all the
kinematical and density effects are fully taken into account.

\begin{figure*}
\hspace{-0.7cm}
\centering
\includegraphics[width = 17cm,viewport=38 50 431 302]{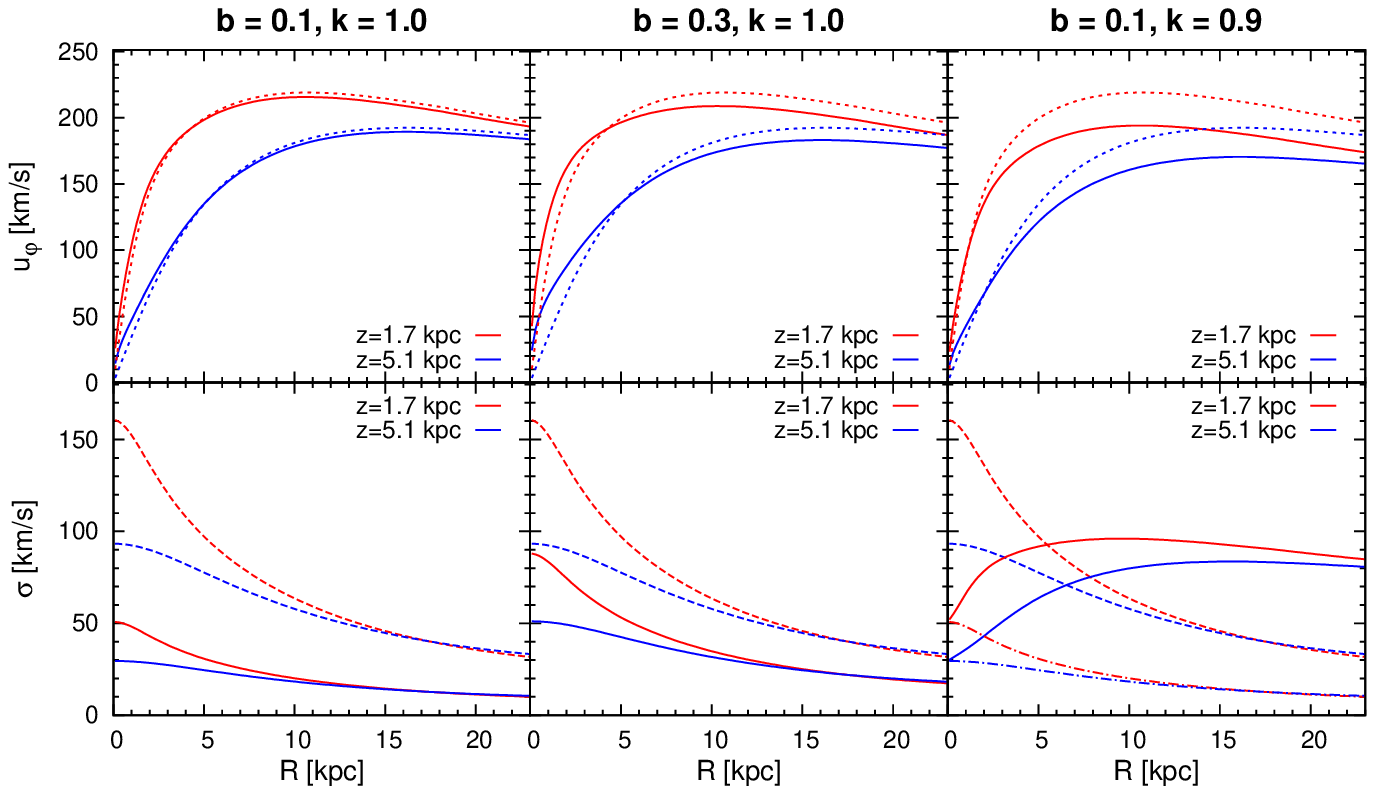}
\caption{Intrinsic rotational velocities and velocity dispersion for
  the anisotropic model $b = 0.1$ and $k = 1$ (left column), the model
  $b = 0.3$ and $k = 1$ (central column), and the model $b = 0.1$ and
  $k = 0.9$ (right column) at $z = 1.7$ kpc and
 $z = 5.1$ kpc. Rotational velocity panels: the dotted lines
  represent the circular velocity $u_{\rm c}$ whilst the solid lines
  are the rotational velocities $u_{\varphi}$. Dispersion panels: the
  dashed lines indicate $\sigma_{z}$, the solid lines
  $\sigma_{\varphi}$ and the dashed-dotted lines $\sigma_{R}$; in the
  first two cases $\sigma_{\varphi} = \sigma_{R}$.}
\label{fig:aniso}
\end{figure*}

Figure \ref{fig:isocube} (first column from the left) shows the pseudo-data cube
constructed from this Jeans model, while each panel in the rightmost column
represents the observed emission of \ion{H}{I} gas in a particular
velocity channel.  The numbers on the top right corners of the data 
column represent the deviations of the line-of-sight
velocities from the systemic one ($528 \, {\rm km} \, {\rm s^{-1}}$),
whilst the shaded area on each map delimits the emission expected from
the disc ($\pm 3 \sigma$ spatial resolution), which we are not
attempting to reproduce with this model. We can see that the observed
\ion{H}{I} emission at higher rotational velocity (corresponding to
the channel at $v_{\rm los}=-236.0 \kms$) is fully due to the
mid-plane, i.e. the halo is not visible.  This is not surprising
because the halo gas rotates more slowly than the gas in the disc and
in fact, as we move towards the bottom panels, the halo emission
becomes important.  Due to its high velocity dispersion, in the
isotropic Jeans model some emission in the halo region is present also
in the upper two velocity channels; only in the middle channel the
agreement is good whilst in the bottom channels the emission is too
horizontally elongated.

Summarizing, it is possible to construct an isotropic Jeans model with
a realistic halo density distribution, in which a negative
vertical gradient in the rotational velocity is present.  However, the
predicted velocity dispersions along the line-of-sight are much higher
than those observed, so that a number of features of the data cube are
not reproduced.  In order to solve these problems, in the next Section
we consider different models with anisotropic velocity dispersion
tensors.

\subsection{Anisotropic models}

The assumption of anisotropy in the velocity dispersion can
overcome, at least in principle, the problems originated by the high
values of $\sigma$ obtained in the isotropic models. 
In Section \ref{sect:twoint} we focus on the classical Satoh
decomposition in a two-integral distribution function (i.e.,
$\sigma_{R} = \sigma_{z}$ everywhere), and in Section
\ref{sect:threeint} we discuss the case of a three-integral
phase-space distribution function for the cloud system. In the following, we indicate with $\sigma_{\rm los}$
the component of the velocity dispersion along the line-of-sight on each place in the galaxy.

\subsubsection{Two-integral anisotropic systems}
\label{sect:twoint}

We examine here cloud systems described by a two-integral
phase-space distribution function, for which $\sigma_{z} =
\sigma_{R}$. The relevant equations are obtained by taking $b = 1$ in
eqs. (\ref{anisojeans}), and adopting a Satoh decomposition in the azimuthal direction.  When $k = 0$
$\sigma_{\varphi}^{2}$ is maximal, while when $k = 1$ the system is
isotropic.  Allowing for $ k = k(R, z)$, more rotationally supported
models can be constructed, up to the maximally rotating one where
$\sigma_{\varphi}^{2} = 0$ (Ciotti \& Pellegrini 1996). First of
all, we investigate a model with constant $k$ and we assume $k = 0.9$; for lower values the
disagreement with the data increases.  We expect that the model is not
better than the isotropic ($k = 1$) case in reproducing the
observations, because $\sigma_{\rm los}$ is increased by the adopted
anisotropy; in particular, as can be inferred from
eq. (\ref{satohdecomp2}), $\sigma_{\varphi} \geq \sigma_{z} =
\sigma_{R}$.  The central column of Fig. \ref{fig:iso} (bottom panel)
clearly illustrates this fact, showing a difference among
$\sigma_{\varphi}$ and the other two components of approximately $30
\, {\rm km\, s^{-1}}$, on average. For what concerns the properties of
the rotation curves, the considerations carried out in the isotropic
case are still valid, because from eqs. (\ref{anisovelocity2}) and
(\ref{satohdecomp}) it follows that $u_{\varphi} = k\,u_{\rm iso}$ so
that in the external regions of the galaxy $u_{\varphi} < u_{\rm c}$.
The pseudo-data cube of the model presented in Fig. \ref{fig:isocube}
(second column) shows that in the channels close to the systemic
velocity (bottom) the emission of this model is even more elongated than in the isotropic
case, in complete disagreement with the data.

\begin{figure*}
\includegraphics[width = 17 cm,viewport=19 121 580 763]{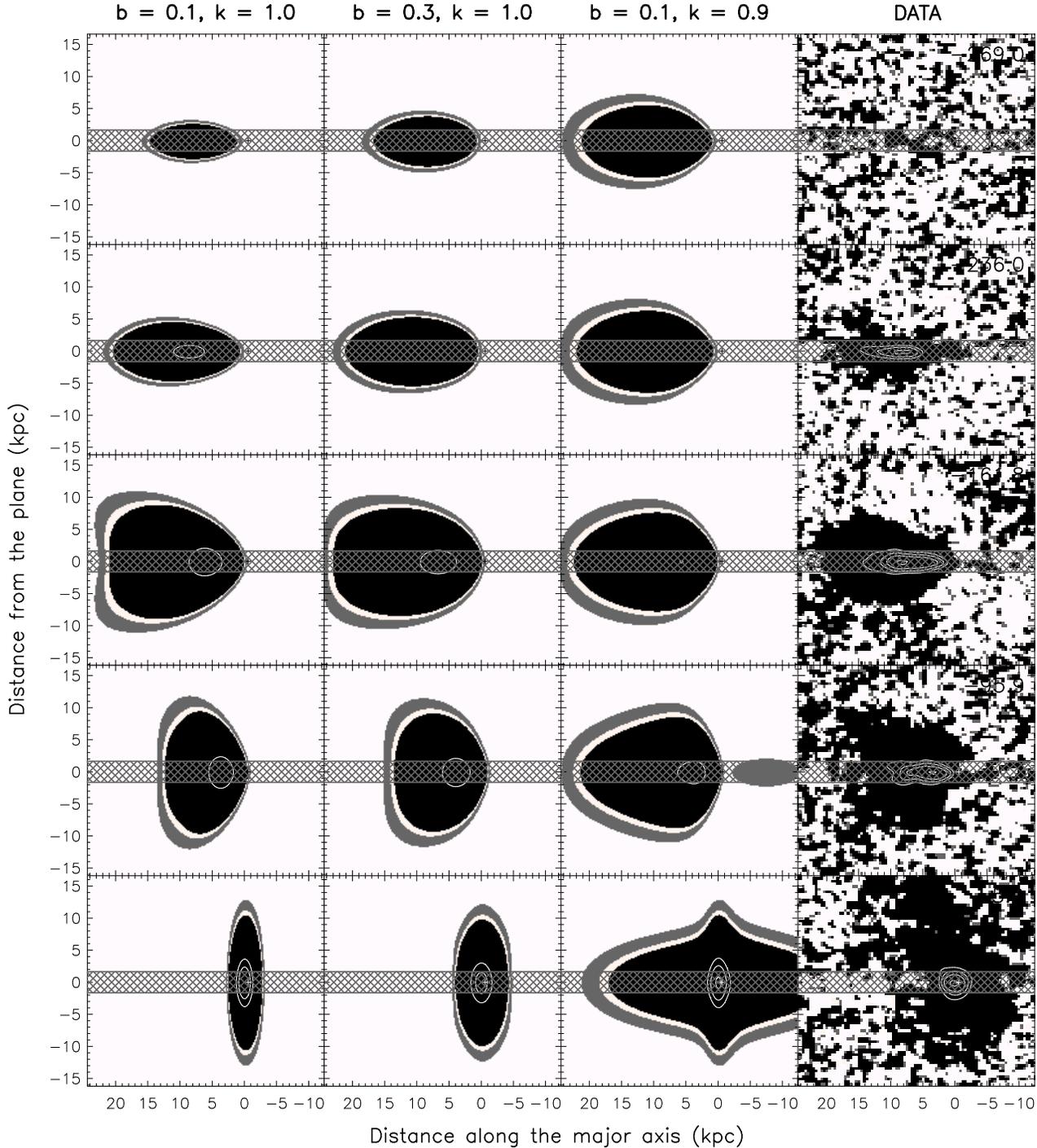}
\caption{Comparison of five representative velocity channels of the
  \ion{H}{I} data cube of NGC 891 and the predictions of 3
  three-integral anisotropic models: $b = 0.1$ and $k = 1$ (first
    column), $b = 0.3$ and $k = 1$ (second column), $b = 0.1$ and $k =
    0.9$ (third column). The shaded area in each channel delimits the
  emission expected from the disc and the contour levels are the same
  as in Fig. \ref{fig:isocube}.}
\label{fig:anisocube}
\end{figure*}

In two-integral systems, the line-of-sight velocity dispersion can be reduced only if
$k > 1$. As a limit case, we present the results for the maximally
rotating model in which $k=k_{\rm max}(R,z) \geq 1$ such that the
rotational velocity has its maximum value everywhere and
$\sigma_{\varphi} = 0$. The corresponding intrinsic rotation curves
are presented in Fig.~\ref{fig:iso} (top right panel)
and their peculiarity is the steep rise in the central regions.
The reduced value of $\sigma_{\rm los}$ improves the agreement with
the data, as it can be seen from the pseudo-data cube of the model in
Fig. \ref{fig:isocube}, especially in the channels near the systemic
velocity.  This agreement is not so good in the channel at $v_{\rm
los}=-236.0 \kms$ and in general in all the channels between $-236.0
\kms$ and $-161.8 \kms$ (not shown here) where, due to the steep rise
in the rotational velocity at small $R$, there is emission in the
central halo regions not seen in the data.

\subsubsection{Three-integral anisotropic systems}
\label{sect:threeint}

We now turn the attention to the models with $b \leq 1$ (therefore
supported by a three-integral phase-space distribution function) and
we first discuss the case $\sigma_{R} = \sigma_{\varphi}$ (i.e. $k =
1$).  The best fit to the data is obtained by an anisotropic model
with $b \sim 0.1$, which limits the line-of-sight velocity dispersion
to about $20 \kms$ for a large interval of $R$ (Fig. \ref{fig:aniso},
bottom left panel).
The rotation curves for this model (top left panel) are very similar to the
circular velocity curves, in accordance with eq. (\ref{anisovelocity2}) 
which states that, in the limiting case $b = 0$, the curves coincide. 
The pseudo-data cube of the model is presented in Fig. \ref{fig:anisocube}
(leftmost column).
In the high velocity channels (top panels)
the halo emission is still present but, at variance with the isotropic
case, this is due to the fact that the rotational velocities of the
model are slightly higher than those observed.
In the low velocity channels (bottom panels) the halo emission is narrower 
than the data along the horizontal axis, due to a rapid decrease of the 
velocity dispersion in the outer parts.
We discuss these discrepancies more in detail in Section \ref{sect:discussion}.

An important issue regarding the anisotropic models is the effect that
the choice of the parameters $b$ and $k$ may have on the properties of
the solutions. We analyse this aspect by fixing the value of one of
the two parameters and by varying the other. The results of this
comparison, for representative values of $b$ and $k$, are shown in
Fig. \ref{fig:aniso}, where the model with $b = 0.1$ and $k = 1$ 
\rev{(which, hereafter, we will also refer to as the ``high-$\sigma_z$ model'')}
has been taken as a reference case. In particular, increasing the value of
$b$ at fixed $k$ has the net result of increasing the line-of-sight
velocity dispersion (i.e. $\sigma_{R}$ and $\sigma_{\varphi}$) whilst
the rotational velocity is decreased; the decrease in rotational
velocity is also obtained when the value of $k$ is decreased at fixed
$b$, but in this case only the value of $\sigma_{\varphi}$ is
affected.  In Fig. \ref{fig:anisocube} we present the pseudo-data
cubes of the models just described.  The emission of the reference
model ($b=0.1$, $k=1$) in all the velocity channels is less
horizontally extended than that of the model with $b=0.3$, due to the
smaller values of $\sigma_{\rm los}$.  Although the $b=0.3$ model
seems to agree better with the data in the channel maps close to
systemic velocity the top panels of Fig. \ref{fig:anisocube} show
substantial halo emission, due to the now increased line-of-sight
velocity dispersion.  The interpretation of the differences between
the reference model and the model with $k \neq 1$ is complicated by
the fact that $\sigma_{\varphi} \neq \sigma_{R}$ so that the extension
of the emission strongly depends on their relative contribution to the
velocity dispersion along the line-of-sight.  In particular, the
importance of $\sigma_{\varphi}$ is maximum in the high velocity
channels (top panels), and then it progressively diminishes in the
channels near the systemic velocity in favour of $\sigma_{R}$.

In short, three-integral models with $b<1$ and $k=1$ perform better
than the isotropic and the two-integral models because $\sigma_{\rm
  los}$ can be reduced to values comparable to the the observed one.
However, a number of features of the data cubes are still not
reproduced revealing intrinsic problems with all these stationary
models that we discuss more in detail in the next Sections.

\begin{figure*}
\centering
\includegraphics[width = 16 cm, height=18cm,viewport=33 92 500 730]{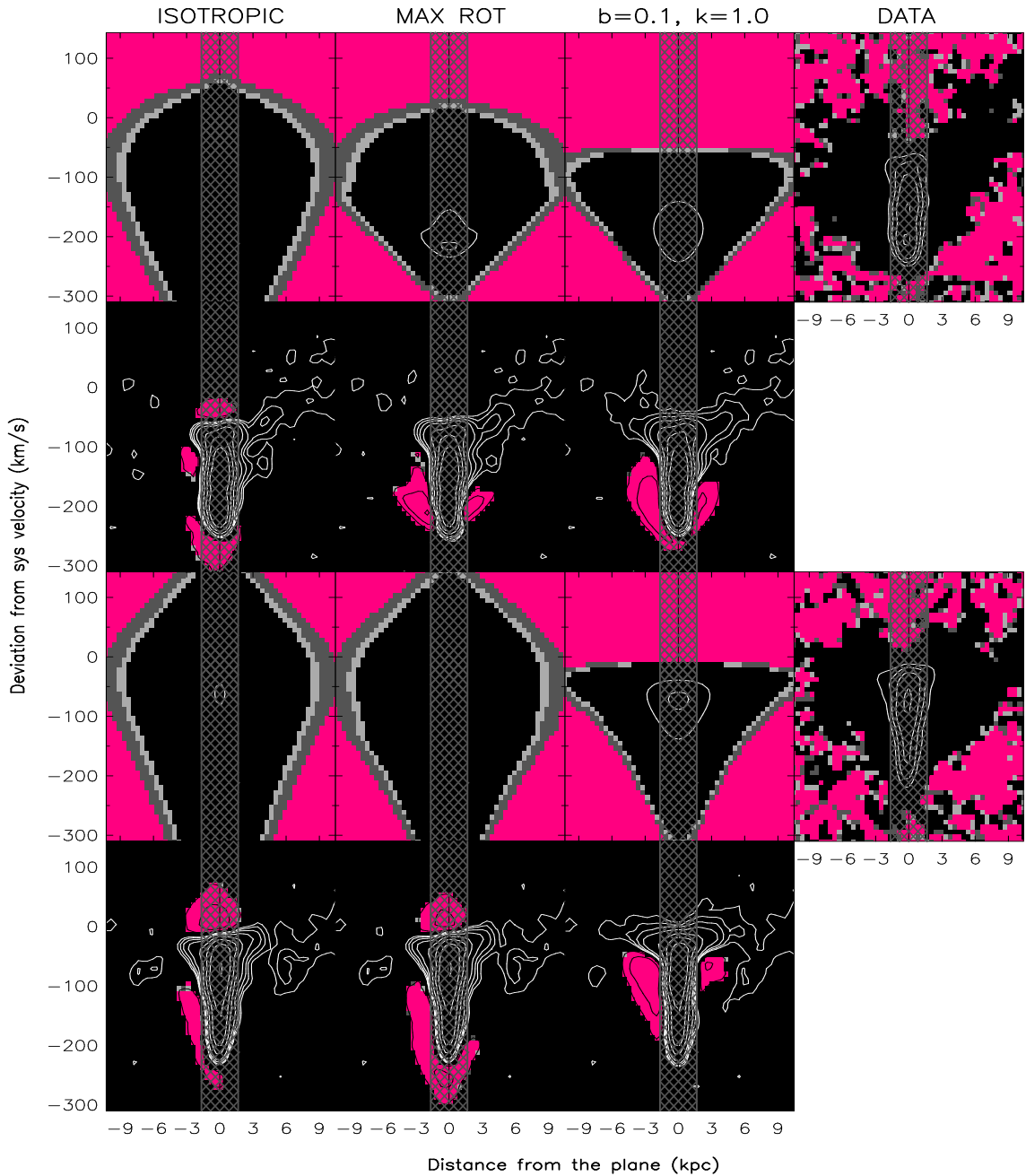}
\caption{Comparison between two position-velocity cuts perpendicular
  to the plane of NGC 891 and some representative models. In the first
  row the cut is made at $R = 7.5$ kpc on side of the galaxy where the
   gas is moving towards the observer, and the corresponding
    residuals, are presented in the second row. In the third row the
  cut is on the same side of the galaxy but at $R = 2.8$ kpc and the
  residuals are shown in the fourth row. The shaded area in each plot
  delimits the emission expected from the disc and the contour levels
  are the same as in Fig. \ref{fig:isocube}. {For the residuals,
    computed subtracting the model to the data, the black contours
    represents positive values whilst the white contours are the
    negative ones; also in this case the contour levels (absolute value) are the same as in Fig. \ref{fig:isocube}.}}
\label{fig:v-z_plot}
\end{figure*}

\subsection{Position-velocity plots}
\label{sect:cuts}

Among the models presented above we select
the three most representative and compare them further with the
\ion{H}{I} data.  Figure \ref{fig:v-z_plot} shows two cuts of data cube
of NGC\,891 (rightmost column) taken perpendicularly to the plane of
the galaxy at $R = 7.5$ kpc (first row) and $R = 2.8$ kpc (third row)
from the centre, towards North-East.  These plots are analogous to
those shown in Fig.\ 16 of Oosterloo et al.\ (2007).  The
position-velocity diagrams are compared with the prediction of three
models: the isotropic model, the maximum rotation model and the
three-integral \rev{high-$\sigma_z$} model.  Also the corresponding residuals
are shown (second and fourth rows) computed as the difference between
the data and the models; the shaded area in each position-velocity
plot delimits the emission expected from the disc.

The characteristic triangular shape of the data diagrams (rightmost column) 
is due to the negative vertical gradient in the rotational velocity of the halo
and is the key feature that the models should reproduce. 
Clearly, this is not the case for the isotropic model
as the high velocity dispersion makes the emission very
elongated, although a negative vertical gradient in the rotational
velocity is visible. 
In the $2.8$ kpc cut (third row), the emission is
so elongated that a substantial amount of gas is apparently ``counter-rotating''
(upper part of the plot).
The maximally rotating model (second column) also does not reproduce the
shape of the emission especially at $R=2.8$ kpc from the centre of the galaxy
(third row).
Also in this case the predicted line-of sight velocity dispersion is too high.
The position-velocity diagrams for the \rev{high-$\sigma_z$} model
(third column) show a qualitative agreement with the data as the 
triangular shape is reproduced.
However the velocity dispersion in the upper halo $|z|> 5$ kpc is clearly not
matched and there is a downward shift of the emission especially visible 
in the $7.5$ kpc cut.
These features are also apparent in the residual maps and in 
particular the \rev{high-$\sigma_z$} model, despite reproducing well
the shape of the data, shows substantial residuals.
In addition, the residuals also reveal the intrinsic asymmetric distribution 
of the extra-planar gas with respect to the mid-plane of NGC 891.

Figure \ref{fig:v-z_plot} gives a different view of the discrepancies
already noted in Figs.\ \ref{fig:isocube} and \ref{fig:anisocube}.
The advantage of a cut perpendicular to the plane of the galaxy is
that it captures in one diagram the kinematics of the extra-planar gas
at different heights and, as mentioned, its characteristic triangular
shape is the evidence of the gradient in rotational velocity with
height.  The only model that reproduces at least qualitatively this
shape is the anisotropic $b=0.1, k=1$ model.

\section{Discussion}
\label{sect:discussion}

In this paper we have explored an alternative interpretation of the baroclinic fluid
homogeneous models for extra-planar gas, i.e. that of a ``gas'' of
\ion{H}{I} clouds described by the stationary Jeans equations.  Due to
the formal equivalence of the equations describing the two systems, a
similarity in the behaviour of the isotropic Jeans model and that of
the baroclinic solutions of Barnab\`{e} et al.\ (2006) is expected.
The present investigation is not just a re-interpretation of the
results of Barnab\`{e} et al.\ (2006): even in the isotropic case our
models differ from theirs in the choice of galaxy mass and cloud
density distributions.  We find that the choice of the mass model has
a little impact on the properties of the solutions, but the form of
the density distribution is critical.  Barnab\`{e} et al. (2006)
adopted a very flat density distribution, in order to reproduce the
off-plane rotation curves.  A flat density distribution requires less
pressure support against the vertical gravitational field of the
galaxy which results in a low temperature of the gas, typically
$10^{5}$ K.  For such a temperature the expected velocity dispersion
in a Jeans isotropic model would have been of the order of $30\,
\rm{km\, s^{-1}}$, close to the mean observed value (Oosterloo et al.\
2007).  Instead, in the present model the parameters of the density
distribution are tuned to reproduce the observed \ion{H}{I}
distribution which has a greater scale-height than that used by
Barnab\`{e} et al. (2006, see eq. [\ref{gasheight}]).  Thus, the
typical values of the velocity dispersion necessary to sustain the
halo are of a factor $\sim\,$3 higher than the above.  As we have
seen, this high velocity dispersion has a large impact on the observable kinematics
of the extra-planar gas and it is the main reason why the isotropic
model fails to reproduce the data.  The higher values of the velocity
dispersion also influence the rotational velocities: indeed the
rotation curves of our Jeans model are about $20\, \rm{km\, s^{-1}}$
lower than those obtained by Barnab\`{e} et al. (2006).

\rev{
The high values of the velocity dispersion of the \ion{H}{I} halo
clouds may appear, at first sight, unphysical.
However we notice that these velocities are unavoidable in any
model attempting to reproduce gaseous halos with 
such large scale-heights (e.g.\ Fraternali \& Binney 2006).
The source of kinetic energy for these clouds must come from
supernovae, indeed hydrodynamical simulations of superbubble expansion
in galactic discs predict these kinds of blowout velocities 
(e.g. Mac Low \& McCray 1988). 
Finally, we also notice that velocities of the order of $100 \kms$ are 
in fact typical for halo
clouds in the Milky Way (Wakker \& van Woerden 1997) and have been observed 
in external galaxies whenever the inclination of the disc along the 
line-of-sight is favorable 
(e.g.\ van der Hulst \& Sancisi 1988; Boomsma et al.\ 2008).
}

As already mentioned in Sect. \ref{sect:cuts}, the most satisfactory model is the anisotropic $b = 0.1$, $k = 1$ model
\rev{(or high-$\sigma_z$ model)}, which has been constructed in order to have $\sigma_{R} =
\sigma_{\varphi} < \sigma_{z}$.  These requirements are roughly in
agreement with the prescriptions for a galactic fountain where the
material is ejected mostly vertically from the disc and falls back
also vertically.  Fraternali \& Binney (2006, 2008) have build
galactic fountain models for NGC\,891 and compared them with the data
cube in the same way we did here.  They found that the halo gas
distribution is well reproduced by fountain clouds that are kicked out
of the disc with a Gaussian distribution of initial velocities and
dispersion of about $80 \kms$, which is a value very similar to the
average $ \sigma_{z}$ predicted by our stationary models (Fig.\
\ref{fig:aniso}).  Moreover, Fraternali \& Binney (2006) investigated
the opening angles about the normal to the plane of the ejected clouds
finding that it has to be lower than 15 degrees, in other words the
clouds are ejected almost vertically.  Therefore, the $b=0.1, k=1$
model is quantitatively similar to the fountain models. 
What makes the difference
between the two is the lack of non-circular (inflow and outflow)
\rev{ordered} motions in our stationary model that, instead, in fountain models are
always present.

The most problematic discrepancy between the \rev{high-$\sigma_z$} model and
the data is visible in the channel map close to the systemic velocity
(see channel at $-5.2 \kms$ in Fig.\ \ref{fig:anisocube}).  This map
shows the gas that has a null velocity component along the
line-of-sight (once the systemic velocity has been removed).  Much of
this gas is located in the outer parts (in $R$) of the galaxy both in
the disc and in the halo.  The width of the channel (along the
horizontal axis) is mostly due to the velocity dispersion of this gas.
The \rev{high-$\sigma_z$} model fails to reproduce the data in two ways: 1) the
channel map is too narrow, 2) the width (in the $x$-axis) of the
emission decreases with height contrary to what seen in the data.
These differences are caused by the fact that the velocity dispersion
in the model decreases both with $R$ and with $z$, whilst in the data the
two behaviours seem reverse.  The fact that the velocity dispersion
increases with height has been clearly shown by Oosterloo et al.\
(2007) who built artificial cubes with different parameters and
compared them with the NGC\,891 data cubes.  These authors also kept
the dispersion constant with $R$ and reproduced much better the shape
of the channel maps close to the systemic velocity (see their Fig.\
14).  The decrease of the velocity dispersion with $R$ and $z$ is a
common problem of all our stationary models and may be related to
the lack of \rev{(ordered)} non-circular motions mentioned above.

\begin{figure}
\includegraphics[width=0.45\textwidth,height=11cm,viewport=50 50 230 302]{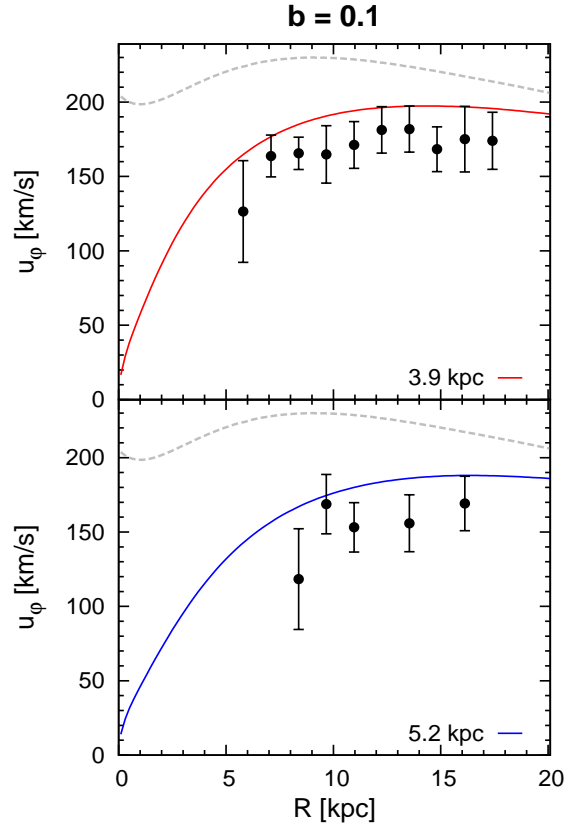}
\caption{Rotational velocities of the model $b = 0.1$ and $k = 1$
    (solid line) and rotation curves in the \ion{H}{I} by Fraternali
    et al. (2005) (points with error bars) at $z = 3.9$ kpc (top) and
    $z = 5.2$ kpc (bottom). For each plot the long-dashed line
    represents the rotational velocity on the mid-plane.}
\label{fig:rotcurve}
\end{figure}

In the \rev{high-$\sigma_z$} model the rotational velocities can be
directly compared with those obtained from the data because the
line-of-sight velocity dispersion has values comparable with the
observed one. In Fig. \ref{fig:rotcurve} we show the rotation curves
of NGC\,891 at distances from the plane of 3.9 kpc (top panel) and 5.2
kpc (from Fraternali et al.\ 2005).  The rotational velocities
predicted by the $b = 0.1, k = 1$ model are shown as solid lines, the
dashed line shows the rotation curve in the plane. In the derivation
of these rotation curves from the data, it has been assumed a constant
value of the velocity dispersion\footnote{Fraternali et al.\ (2005)
assumed $\sigma_{\rm gas} \sim 8\, \rm{km\, s^{-1}}$, for the disc and
the halo.  Oosterloo et al.\ (2007) found that the velocity dispersion
in the halo is actually larger and about $20\, \rm{km\, s^{-1}}$, thus
it is appropriate to shift the data points of Fraternali et al.\
(2005) down by $11\, \rm{km\, s^{-1}}$ (see Fraternali \& Binney
2008 for details).  This is equivalent to derive the rotation curves
assuming a constant $\sigma_{\rm gas} \sim 20\, \rm{km\, s^{-1}}$ for
the halo region.} ($20\, \rm{km\, s^{-1}}$).  The curves have been
derived using the envelope tracing method, which is such that the
larger is the velocity dispersion one assumes the lower is the
obtained rotational velocity.  In our \rev{high-$\sigma_z$} model the
velocity dispersion is not constant along $R$ and it is exactly $20
\kms$ only at $\sim 10$ kpc (Fig.\ \ref{fig:aniso}).  Ideally one
should correct the data points for this effect obtaining slightly
lower velocities for $R<10$ kpc and higher for $R>10$ kpc but the
effect is minimal.  Overall, for what concerns the rotation curves of
NGC\,891 we can say that the \rev{high-$\sigma_z$} stationary model predicts
a gradient that is quite close to (although slightly lower than) the
observed one.

The stationary models presented here do not aim to explain the origin
of the extra-planar gas.  Instead their goal is to find a stationary
description for the extra-planar gas phenomenon whatever physical
mechanism is producing it, assuming that it can be considered roughly
stationary.  This assumption should be reasonably valid if the
extra-planar gas is produced for instance by a galactic fountain and
the star formation rate of the galaxy does not vary considerably with
time as it seems to be the case for galaxies like the Milky Way (e.g.\
Twarog 1980).  The investigation presented by Fraternali \& Binney
(2006) shows that the rotational velocities predicted by a galactic
fountain model are systematically higher than the one derived from the
data and also higher than those predicted by the stationary \rev{high-$\sigma_z$}
model shown in Fig.\ \ref{fig:rotcurve}.  This latter result is
not surprising given that in the fountain model the gas clouds move
outwards for a large fraction of their trajectories and thus tend to
be at ${\rm v}_{\varphi} > u_{\rm c}$ for most of their time ($u_{\rm
c}$ is the circular speed at that position).  On the contrary, in the
\rev{high-$\sigma_z$} model the gas has basically ${ u}_{\varphi} \simeq u_{\rm
c}$ (Fig.\ \ref{fig:aniso}).  Fraternali \& Binney (2008) revised the
fountain model by including interactions of the fountain clouds with
the surrounding ambient medium (see also Barnab\`e et al. 2006 for a similar suggestion).
These interactions cause the fountain
clouds to lose part of their angular momentum and to decrease
their rotation velocities.  Such a model is to date the best
description for the dynamics of the extra-planar gas.  In
addition to the rotational velocities, this latter model reproduces
the data cube in detail including the channel maps close to the
systemic velocity (see Fig.\ 2 of Fraternali \& Binney 2008). Kaufmann
et al. (2006) also reproduced the rotation curves of NGC\,891 fairly
well with \rev{a model in which the extraplanar gas forms
by condensation of the extended warm-hot corona,} but they did not attempt to
build pseudo-data cubes \rev{to compare directly with the observations}.
\rev{However, the thermal instability
on which such a model is based is unlikely to occur in galactic coronae, because
of the combined effect of buoyancy and thermal conduction
(Binney, Nipoti \& Fraternali 2009).} 

Whatever the origin of the extra-planar gas, it is important to
investigate whether or not it is in a stationary state.  If so, it
should be possible to find stationary models that reproduce these data
in detail.  This paper has been the second attempt of this sort after
Barnab\'{e} et al. (2006).  In both these attempts non-circular motions of the gas
clouds have been neglected for simplicity, but in
the kinematics of the extra-planar gas this kind of motions are clearly present
(e.g.\ Boomsma et al. 2008; Fraternali et al. 2001) and they
could affect substantially the \rev{observed kinematics} of the halo.
\rev{
More likely, the cause of the discrepancy between our treatment and
the observations lies on the assumptions of a Jeans-based model
and in particular on the ballistic nature of the cloud motion.
Not only the drag between the \ion{H}{I} halo clouds and the corona may
have an important dynamical effect but also 
at the turbulent cloud/corona interface there may be a considerable
exchange of mass.
Modelling this interface is a very challenging numerical
problem (e.g. Vieser \& Hensler 2007 a,b) and a secure estimate of the rate of mass exchange 
is not yet available.
This issue will be studied in a forthcoming paper (Marinacci et al., in prep.).
}

\section{Summary and conclusions}

In this paper, motivated by the results of Barnab\`e et al. (2006), we
investigated the possibility that the extra-planar gas in spiral
galaxies can be modelled as a ``gas'' composed by cold \ion{H}{I}
clouds that follows the stationary Jeans equations. 
\rev{In doing this we are assuming that the pressure-bound clouds are
moving almost ballistically in the halo.
This assumption is valid in the limit of massive clouds having
negligible rates of mass exchange with the corona.
}

In this alternative interpretation of fluid baroclinic models the thermal
pressure is replaced by an isotropic velocity dispersion tensor, so
that the problem of high temperature of gaseous homogeneous models
is eliminated. We have also extended the discussion to simple
phenomenological models with anisotropic velocity dispersion tensors.
We constructed both isotropic and anisotropic models in the
well-constrained gravitational field of the spiral galaxy NGC 891, and
compared their predictions to the observed kinematics of the
extra-planar gas in that galaxy.  For each model we built a
pseudo-data cube with the same resolution and total flux as the
observations.  The main results of our analysis can be summarized as
follows:

   \begin{enumerate}

   \item The adopted functional form of the cloud density distribution, 
         taken by the observations, leads to physically acceptable solutions in all the models
     investigated. The cloud density distribution is centrally
       depressed, and, in order to match the vertical extension of the
       \ion{H}{I} halo of NGC 891, it has higher scale-height than that used by
       Barnab\`{e} et al. (2006).

   \item All the models computed show a negative vertical gradient in
     the rotational velocity, the distinctive feature of the
     kinematics of the extra-planar gas. 

      \item The support against the vertical gravitational field of
      the galaxy requires a $\sigma_{z} \simeq 50 - 100\, \rm{km\,
      s^{-1}}$ and therefore the line-of-sight velocity dispersion of
      the isotropic model is a factor $\sim\,3-4$ higher than that 
      observed.

    \item With the introduction of the anisotropy it is possible
        to restrict the line-of-sight velocity dispersion to the
        observed values, but the predicted vertical gradient in the
      rotational velocity is somewhat too shallow and other features
      of the data cube are not fully reproduced.

   \end{enumerate}

We conclude that the dynamics of extra-planar gas in a galaxy like
NGC\,891 is not fully described by any of the stationary models
considered here.  However a model with an anisotropic velocity
dispersion tensor, which mimics a galactic fountain is the
preferable among all. \rev{The fact that none of the stationary Jeans
models analysed here can reproduce all the features of the observed
(extra-planar) gas kinematics might suggest that
the cloud motion is not purely ballistic, and that the interaction between
the clouds and the coronal gas, perhaps in the form of mass exchange, 
plays an important dynamical role.}


\section*{Acknowledgments}
We thank G. Bertin for helpful discussions and J.J. Binney for 
reading the manuscript and providing precious suggestions to 
improve the presentation of the paper. We also thank 
an anonymous referee for his/her valuable comments.


\bsp

\label{lastpage}

\end{document}